\documentclass[pdftex]{article}

\usepackage{PRIMEarxiv}

\usepackage[utf8]{inputenc} 
\usepackage[T1]{fontenc}    
\usepackage{hyperref}       
\usepackage{url}            
\usepackage{booktabs}       
\usepackage{amsfonts}       
\usepackage{nicefrac}       
\usepackage{microtype}      
\usepackage{lipsum}
\usepackage{fancyhdr}       
\usepackage{graphicx}       
\usepackage{amsmath,amssymb,amsfonts,mathtools}
\usepackage{multirow}

\newcommand{\argmin}{\mathop{\rm arg~min}\limits}
\graphicspath{{media/}}     

\title{Exploring internal representation of self-supervised networks: few-shot learning abilities and comparison with human semantics and recognition of objects}

\author{
  Asaki Kataoka \\
  Graduate School of Arts and Sciences \\
  The University of Tokyo \\
  Meguro, Tokyo, Japan \\
  \texttt{asaki-kataoka@g.ecc.u-tokyo.ac.jp} \\
   \And
  Yoshihiro Nagano \\
  Graduate School of Informatics \\
  Kyoto University \\
  Sakyo, Kyoto, Japan \\
  \texttt{nagano@i.kyoto-u.ac.jp} \\
   \AND
  Masafumi Oizumi \\
  Graduate School of Arts and Sciences \\
  The University of Tokyo \\
  Meguro, Tokyo, Japan \\
  \texttt{c-oizumi@g.ecc.u-tokyo.ac.jp}
}

\begin{document}
\maketitle

\begin{abstract}
Recent advances in self-supervised learning have attracted significant attention from both machine learning and neuroscience. This is primarily because self-supervised methods do not require annotated supervisory information, making them applicable to training artificial networks without relying on large amounts of curated data, and potentially offering insights into how the brain adapts to its environment in an unsupervised manner. Although several previous studies have elucidated the correspondence between neural representations in deep convolutional neural networks (DCNNs) and biological systems, the extent to which unsupervised or self-supervised learning can explain the human-like acquisition of categorically structured information remains less explored. In this study, we investigate the correspondence between the internal representations of DCNNs trained using a self-supervised contrastive learning algorithm and human semantics and recognition. To this end, we employ a few-shot learning evaluation procedure, which measures the ability of DCNNs to recognize novel concepts from limited exposure, to examine the inter-categorical structure of the learned representations. Two comparative approaches are used to relate the few-shot learning outcomes to human semantics and recognition, with results suggesting that the representations acquired through contrastive learning are well aligned with human cognition. These findings underscore the potential of self-supervised contrastive learning frameworks to model learning mechanisms similar to those of the human brain, particularly in scenarios where explicit supervision is unavailable, such as in human infants prior to language acquisition.
\end{abstract}


\section{Introduction}
Self-supervised learning has recently gained significant attention from both the machine learning and neuroscience communities. Unlike supervised learning, which requires explicit task-specific labels, self-supervised learning relies on inherent structures within the data itself and does not require manual supervision. This property makes it particularly advantageous in machine learning, enabling models to be trained on vast amounts of uncurated (unlabeled) data.
Recent studies have demonstrated the effectiveness of self-supervised learning as a powerful method for representation learning \cite{Arora2019-cu, Medina2020-iy, Chen2020-zq, Newell2020-ff, pmlr-v162-bao22e, Ericsson2021-oq, Nozawa2021-rb, Shi2021-qi, Wang2021-ag, Zhu2022-tp, Hu2024-vz}.

In neuroscience, it is equally important to investigate the characteristics of neural representations that emerge from self-supervised learning, as this can provide insights into learning mechanisms in the brain. Given that self-supervised learning does not require labeled input, it offers a plausible framework for brain-like learning. In particular, since language is considered a major source of supervision in humans \cite{Knudsen1994-yf, Glaser2019-bj, Loewenstein2021-mw}, self-supervised learning may play a central role in the brains of human infants before language acquisition, as well as in non-linguistic animals. When applied to the study of neural learning and information representation, self-supervised learning may help explain empirical findings showing that prelinguistic infants exhibit cognitive abilities--such as categorical representation--similar to those of adults \cite{Carey1978-nc, Quinn1993-ol, Behl-Chadha1996-qx, Freedman2001-md, Smith2002-up, Yang2016-lc}.

Deep convolutional neural networks (DCNNs) have frequently been used as computational models of neural circuits to study such neural representations. Earlier studies have shown that DCNNs trained with supervised learning exhibit representational similarities to the visual systems of humans and animals \cite{Lecun1998-up, Kriegeskorte2008-bz, Jarrett2009-kf, Krizhevsky2012-mo, Yamins2013-pu, Yamins2014-pi, Khaligh-Razavi2014-an, Majaj2015-nf, Yamins2016-mu, Rafegas2018-ak, Rajalingham2018-su, Hebart2020-pv, Marques2021-kx, Kawakita2024-pw}. Building on this foundation, recent work has demonstrated that DCNNs trained using self-supervised algorithms also show representational similarities to biological visual systems \cite{Bakhtiari2021-fw, Zhuang2021-ly, Nayebi2021-nv, Konkle2021-ox, cadena2019how, Konkle2022-wo, Millet2022-kd, Prince2024-yp}, further supporting their plausibility as models of the visual system.

In this study, we investigate the internal representations of deep convolutional neural networks (DCNNs) through the lens of inter-category relationship structures, as revealed by \textit{few-shot learning} performance. Few-shot learning refers to the ability to recognize novel, previously unseen categories using only a limited number of examples. While prior studies have highlighted the similarity between DCNN representations and those of humans and animals, the structure of category-level representations has not been thoroughly explored. A recent study \cite{Sorscher2022-dr} evaluated the few-shot learning capabilities of DCNNs trained with both supervised and self-supervised methods. Building upon this work, we compare the category structures revealed through few-shot learning in DCNNs with human semantic organization and recognition performance, aiming to further clarify the nature of internal representations learned without explicit supervision.

In this paper, we pursue three objectives: (i) to confirm that DCNNs trained with self-supervised learning can perform few-shot learning accurately, and (ii) to investigate their internal representations by comparing them to human semantic organization and (iii) human recognition performance. The few-shot learning ability (i) is evaluated based on the linear separability of categories within the neural representation space. From this evaluation, we derive category-wise confusion matrices for each trained DCNN. These matrices are then used to analyze the inter-category structure of the representations and to compare them to (ii) human semantic structures and (iii) human recognition patterns. While prior studies have investigated self-supervised DCNNs \cite{Medina2020-iy, Sorscher2022-dr, Lu2022-wx}, our novel contribution lies in the comparative analyses involving human-level semantics and recognition—specifically, objectives (ii) and (iii).

Our experimental results indicate that the internal representations arising from self-supervised learning in DCNNs closely resemble human semantic structures and recognition patterns.
These findings suggest that inter-category structures similar to those found in human cognition can emerge even before the application of explicit supervision, thereby supporting both the biological plausibility and practical utility of self-supervised learning in brain-like systems.

\section{Materials and Methods}
\subsection{Overview}
\begin{figure}
    \centering
    \includegraphics[width=\linewidth]{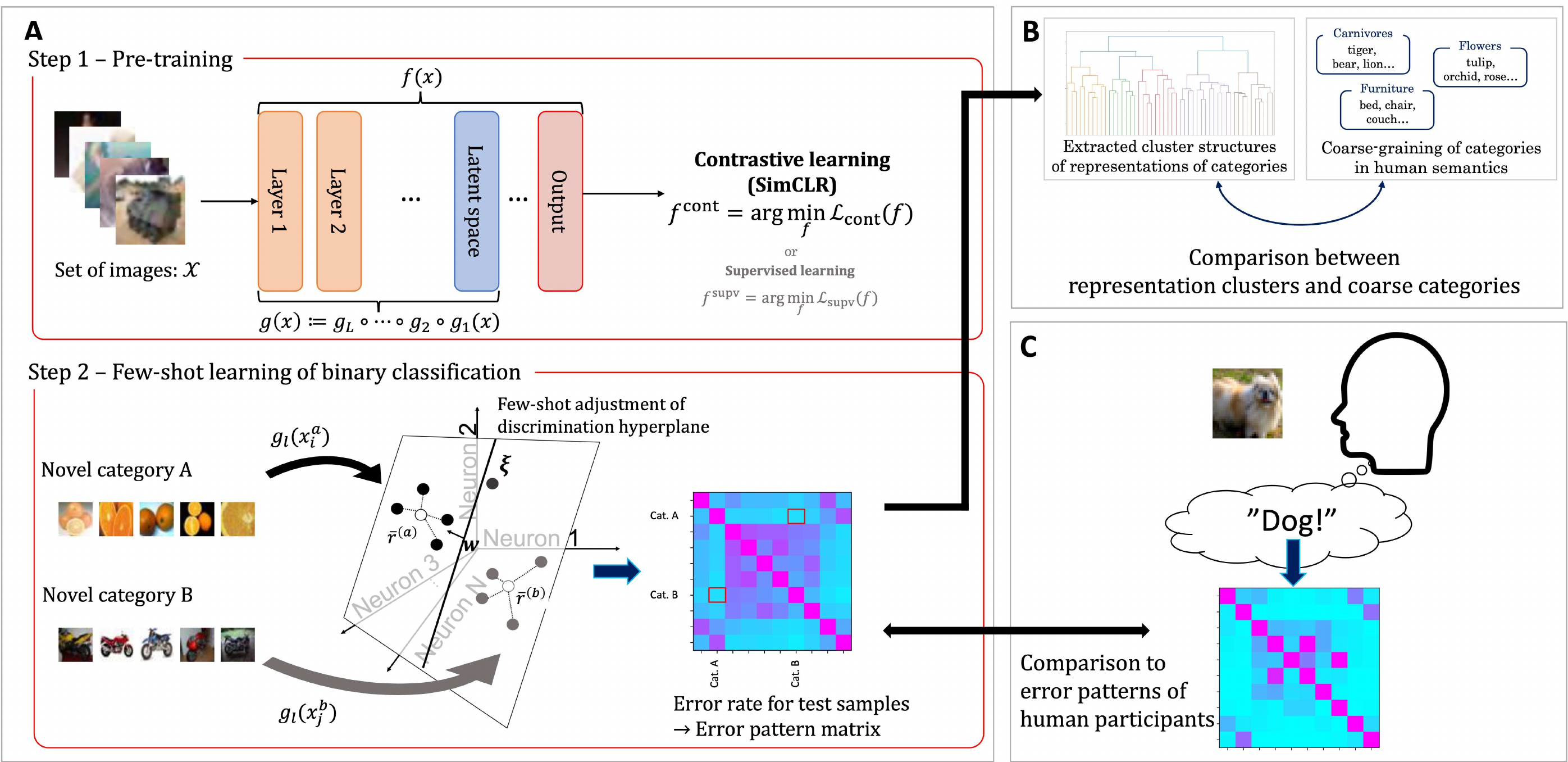}
    \caption{Schematic illustration of the experimental procedure in this study. (A) A two-step methodology for evaluating the few-shot learning performance of DCNNs. In Step 1 (pre-training), the network is trained using self-supervised contrastive learning. In Step 2, pairwise few-shot classification is performed, and performance is assessed using error pattern matrices, where each cell represents the classification error rate between a pair of novel categories. (B) Clusters of object categories derived from the error pattern matrices are compared to coarse-grained category groupings based on human semantic relationships. (C) Similarity is evaluated between the error pattern matrices of the DCNNs and the confusion matrix obtained from human participants performing an object classification task.}
    \label{fig:schematic_method}
\end{figure}

In this study, we adopt a two-step procedure to evaluate the few-shot learning performance of DCNNs trained using a self-supervised learning algorithm (Fig.~\ref{fig:schematic_method}). In the first step, the DCNNs are pre-trained with a self-supervised \textit{contrastive} learning method. To assess the impact of the absence of supervision, we also train a separate DCNN using a supervised object classification task, which serves as a baseline for comparison. The evaluation of few-shot learning performance is conducted in the second step.

In the initial pre-training step (Fig.~\ref{fig:schematic_method}A, top), we train a DCNN using a self-supervised contrastive learning algorithm \cite{Jaiswal2020-dy, Kumar2022-gj}. During this step, the synaptic weight parameters of the network are optimized to minimize an objective function. The loss function is designed such that it does not require explicit supervision signal during its training procedure. In order to investigate the influence of absence of supervision signal, we also train another DCNN using supervised learning framework.

In the second step (Fig.~\ref{fig:schematic_method}A, bottom), we evaluate the few-shot learning performance of the DCNNs, specifically the linear separability of internal representations into the target categories. During this evaluation, the synaptic weights of the DCNNs are kept frozen. The networks are presented with images from novel object categories that were not included during pre-training. A few exemplar images from each category are used to compute ``prototype'' representations, and the remaining samples are classified based on their similarity to these prototypes. The classification results are summarized in confusion matrices, which we refer to hereafter as error pattern matrices.

After obtaining the error pattern matrices, we perform additional analyses to examine how closely the internal representation structures of the networks resemble those of humans. To evaluate the similarity of these structures in detail, we adopt the following two approaches.

The first approach (Fig.~\ref{fig:schematic_method}B) evaluates how the grouping of object categories in the internal representations of DCNNs aligns with human semantic organization. Using the error pattern matrices obtained from the few-shot learning task, we perform hierarchical clustering to identify clusters of categories that are represented closely together. We then quantify the extent to which the categories within each cluster correspond to predefined coarse-grained object categories.

In the second approach (Fig.\ref{fig:schematic_method}C), we quantitatively evaluate the similarity of error patterns between human participants and DCNNs in object classification tasks. Specifically, we use a dataset of object images and a confusion matrix derived from human participants performing multi-label classification of these images (see Section\ref{sec:cifar10h}). We then compare this human confusion matrix with confusion matrices obtained from the multi-class few-shot learning evaluations of the networks.

\subsection{Pre-training}
\begin{figure}
    \centering
    \includegraphics[width=\linewidth]{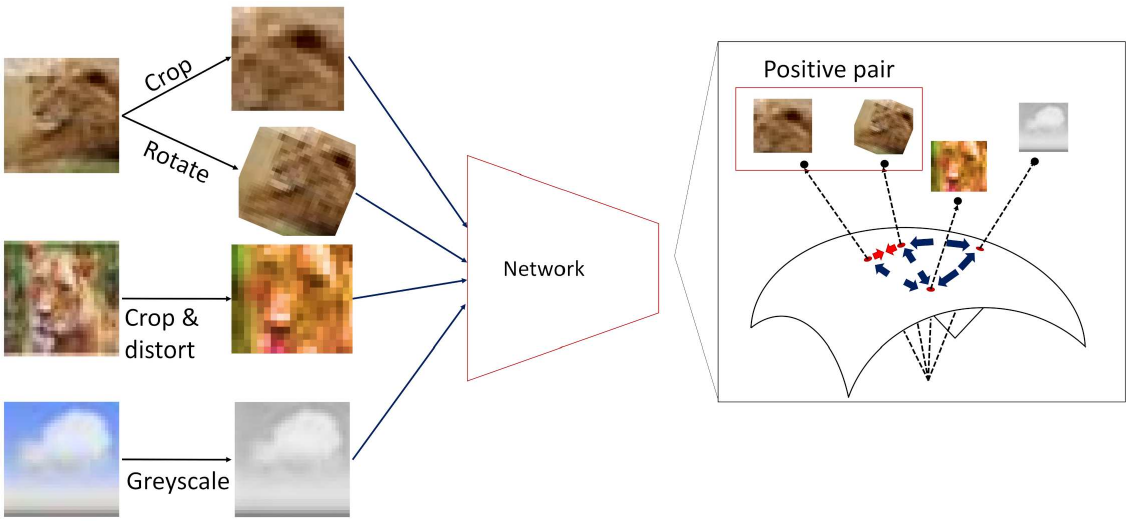}
    \caption{Schematic illustration of SimCLR contrastive learning. During SimCLR training, the DCNN is provided with a large number of image inputs, each generated from an original image by applying random augmentations such as cropping, rotation, or color distortion. In SimCLR, the network is trained so that internal representations of augmented views from the same original image (i.e., \textit{positive pairs}) are mapped close together in the latent space, while representations of all other combinations (\textit{negative pairs}) are pushed farther apart.} 
    \label{fig:schematic_contrastive}
\end{figure}

In the first step, we trained DCNNs $f \colon \mathcal{X} \rightarrow \mathcal{Y}$ using self-supervised contrastive learning, and also trained a separate model with supervised object classification as a baseline. Both models share the same encoder architecture $g \colon \mathcal {X} \rightarrow \mathcal{Z}$, which maps an input image to a common latent space. From this latent space, distinct projection heads $\operatorname{proj} \colon \mathcal {Z} \rightarrow \mathcal {Y}$ are used, such that $f\coloneqq \operatorname{proj} \circ g$. Note that the projection heads differ between the two models, and consequently, the dimensionality of $\mathcal{Y}$ is not the same across them.
\subsubsection{Self-supervised contrastive learning}
In this work, we adopt \textit{self-supervised contrastive learning} as a representative framework of learning rules that do not rely on explicit supervision signals. In contrastive learning, a network is trained such that the internal representations of \textit{semantically similar} inputs (positive pairs) are brought closer together, while those of dissimilar inputs (negative pairs) are pushed apart in the network’s latent space. Specifically, we use SimCLR \cite{Chen2020-ih} as a standard contrastive learning algorithm. We also include SimSiam \cite{Chen2020-zq} as an additional contrastive method to evaluate the robustness of our findings (see Supporting Text and Supplementary Fig. S1).

SimCLR applies a randomly selected combination of augmentations to each input image and treats two differently augmented views of the same image as a positive pair during training. Fig~\ref{fig:schematic_contrastive} provides a schematic illustration of instances of such augmentations, including random cropping, rotation, color distortion, and grayscaling. Table \ref{tab:augmentation_rules} outlines the specific augmentation rules and parameters we used.
\begin{table}[h]
\centering
\begin{tabular}{ll}
\hline
\textbf{Name of augmentation} & \textbf{Parameters} \\ \hline \hline
\multirow{2}{*}{Random cropping} & Scale: [0.08, 1.0] \\
                              & Ratio: [0.75, 1.25] \\ \hline
\multirow{1}{*}{Horizontal flipping} & Probability: 0.5 \\ \hline
\multirow{2}{*}{Color jittering} & Strength: 0.5 \\
                                & Probability of grayscaling: 0.2 \\ \hline
\end{tabular}
\caption{Augmentation rules and corresponding parameters.}
\label{tab:augmentation_rules}
\end{table}

Suppose an augmentation function $a\in\mathcal A$ is sampled from a probability distribution $\rho_\text{aug}\in\mathbb P(\mathcal A)$, and an input image $x\in\mathcal X$ is sampled from another distribution $\rho_\text{im}\in\mathbb{P}(\mathcal X)$. Here, $a \colon \mathcal{X} \rightarrow \mathcal{X}$ represents a single composition of randomly applied augmentations listed in Table \ref{tab:augmentation_rules}. The neural network $f \colon \mathcal{X} \rightarrow \mathbb{R}^d$ is trained on those augmented input samples. The \textit{InfoNCE} loss function for SimCLR is defined as
\begin{align}
    &\mathcal{L}_\text{CLR}(f) \coloneqq -\mathbb{E}_{x, \{x_k^-\}\sim\rho_\text{im}^{K+1},\{a_k\}\sim\rho_\text{aug}^{K+2}} \left[l_\text{CLR}(a_{K+1}(x), a_{K+2}(x), \{a_k(x_k^-)\}_{k=1}^{K}; f)\right], \label{eq:info_nce_1} \\
    &l_\text{CLR} \left(\tilde x, \tilde x^+, \{\tilde x_k^-\}_{k=1}^K; f \right) \coloneqq \log\frac{\exp\left(\hat f(\tilde x)\cdot \hat f(\tilde x^+)\right)}{\exp\left(\hat f(\tilde x)\cdot \hat f(\tilde x^+)\right) + \sum_{k=1}^K\exp\left(\hat f(\tilde x)\cdot \hat f(\tilde x_k^-)\right)}, \label{eq:info_nce_2}
\end{align}
where $\{x_k\}_{k=1}^K \sim \rho^K$ indicates that $\{x_1, \ldots, x_K\}$ are independently sampled from the same distribution $\rho$, and $\hat f(\cdot)\colon=f(\cdot) / \|f(\cdot)\|$ denotes the normalized internal representation. Here, $\tilde{x}$, $\tilde{x}^+$, and $\{\tilde{x}^-_k\}_{k=1}^K$ represent the anchor, the positive, and the negative samples, respectively. As shown in Eq~\ref{eq:info_nce_1}, positive pairs are generated by applying different random augmentations to the same image. In Eq~\ref{eq:info_nce_2}, the symbol $\cdot$ denotes the inner product between two vectors. Minimizing Eq~\ref{eq:info_nce_2} can be interpreted as maximizing similarity of representations within the positive pair $(\tilde x, \tilde x^+)$, while minimizing the similarity of them to the negative samples $\{\tilde x_k^-\}_{k=1}^K$. Note that computing the exact expectation in Eq~\ref{eq:info_nce_1} is computationally infeasible due to  multiple integrals over continuous random augmentations. Therefore, we approximate it using the empirical mean over a minibatch.

\subsubsection{Supervised object classification learning}
For comparison with the network trained using the contrastive learning algorithm, we also consider supervised object classification learning. This approach requires explicit supervision signals that specify the object category to which each input image belongs. The network is trained such that its output, interpretable as estimated probabilities over the object categories, closely matches the ground-truth labels provided by the supervision signals.

The loss function for the network $f \colon \mathcal{X} \rightarrow \mathbb{R}^{|\mathcal{C}|}$ is defined as
\begin{equation}\label{eq:cross-entropy}
    \mathcal{L}_\text{supv}(f)=-\mathbb{E}_{(x, y)\in\mathcal{D}}\sum_{c\in\mathcal{C}}y_c\log \text{softmax}_c(f(x)),
\end{equation}
where $\mathcal D:=\{(x_i, y_i)\}_{i=1}^{N}$ denotes the dataset, $\mathcal C$ is a pre-defined set of object categories in the training datas, and $f$ indicates the network being trained. The subscript $c$ denotes the index of a $|\mathcal{C}|$-dimensional vector. $y$ is an element of a probability simplex $\Delta^{|\mathcal{C}|}$, and is typically a \textit{one-hot} vector, \textit{i.e.}, $\{y\in\{0, 1\}^{|\mathcal C|}|\sum_{c\in\mathcal{C}}y_c=1\} \subset \Delta^{|\mathcal{C}|}$. The softmax function, which outputs the probability that $x$ belongs to category $c$, is defined as
\begin{equation}
    \text{softmax}_c(f(x)) = \frac{\exp\left(f_c(x)\right)}{\sum_{c'\in\mathcal C}\exp\left(f_{c'}(x)\right)}.
\end{equation}
Optimization is performed using gradient descent with error back-propagation \cite{Rumelhart1986-rm}. To improve robustness to noise, we also applied random augmentations to the input images. The set of augmentations was identical to that used in the self-supervised contrastive learning setting (see Table \ref{tab:augmentation_rules}). As in the contrastive learning case, the expectation in Equation~\ref{eq:cross-entropy} is approximated by empirical average over minibatches due to computational constraints.

\subsection{Network architecture}
\begin{figure}
    \centering
    \includegraphics[width=0.6\linewidth]{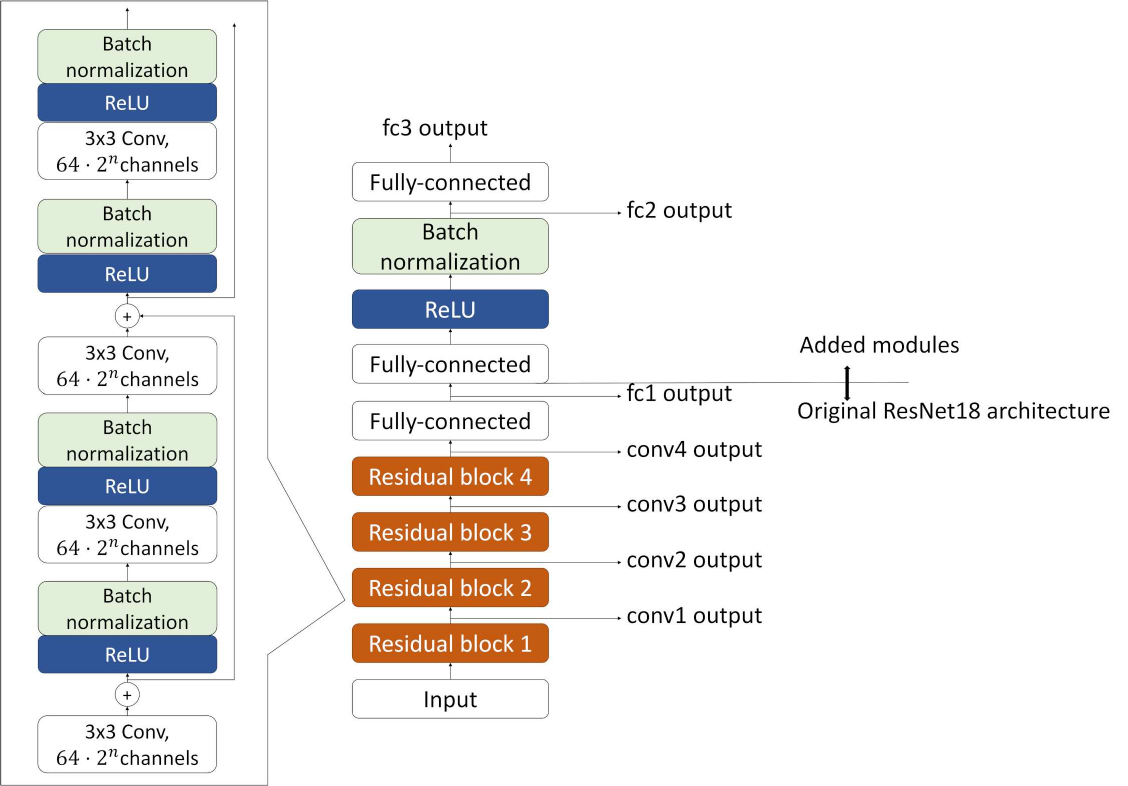}
    \caption{Employed architecture of ResNet-18 network. In addition to the originally proposed architecture, we added additional layers. The objective functions for both contrastive learning and supervised learning are computed in the output of the final layer.}
    \label{fig:resnet18_arch}
\end{figure}

In the present study, we employ a modified ResNet-18 \cite{He2015-vx} as the encoder backbone for the neural networks. The original ResNet-18 architecture consists of 18 layers, including residual connections. To stabilize the learning process, particularly for SimCLR, we added a fully connected layer followed by a ReLU activation, batch normalization, and a second fully connected layer (Fig~\ref{fig:resnet18_arch}). Note that in both the supervised and contrastive learning settings, the function $f$ appearing in the loss definitions refers to the output of this final additional module. Accordingly, the output dimensionality of the final fully connected layer is set to $d$ for contrastive learning and $|\mathcal C|$ for supervised learning settings.

\subsection{Few-shot learning}
One of the main goals of this study is to investigate whether DCNNs trained with self-supervised learning algorithms can accurately perform few-shot classification of novel object categories. To this end, we follow the approach \cite{Sorscher2022-dr} and formalize few-shot learning as the linear separability of internal representations of the novel categories.

The details of the few-shot learning evaluation procedure are as follows. Let $\{c_1, ..., c_n\}$ that have not been used for the pre-training phase. From each category $c_j$, we randomly sample $m$ training examples (with $m=10$ in this study), denoted as $x_i^{(c_j)}\,(i=1,..., m)$. During pre-training, the network output is computed as $f(x)=(\operatorname{proj}\circ g)(x)$, where $g$ is the encoder and $\operatorname{proj}$ is the projection head. In contrast, for few-shot evaluation, we extract representations from the $l$-th layer of the encoder, denoted $r=g_l(x)$, where $g=g_1\circ ... \circ g_L$. For each novel category $c_j$, we compute a ``prototype'' representation by averaging the internal representations of its training samples:
\begin{equation}
    \bar r^{(c_j)}:=m^{-1}\sum_{i=1}^nr_i^{(c_j)}.
\end{equation}
Given these prototypes, a test sample $\xi$ is classified into the category $\tilde c$ defined by
\begin{equation}
    \tilde c=\argmin_j \left\|\bar r^{(c_j)} - \xi\right\|_2.
\end{equation}
This procedure evaluates whether the DCNN organizes internal representations such that inputs from the same category are embedded closely, while those from different categories are well separated.

In particular, when $n=2$, this procedure admits an alternative geometric interpretation. Given the prototypes $\bar r^{(c_1)}$ and $\bar r^{(c_2)}$ for the two categories $c_1$ and $c_2$, we can define a linear decision boundary as follows:
\begin{align}
	&w = \bar r^{(c_1)} - \bar r^{(c_2)}, \\
	&\beta = \frac{1}{2}w\cdot(\bar r^{(c_1)} + \bar r^{(c_2)}).
\end{align}
For a test sample with representation $\xi$, the predicted category is $c_1$ if
\begin{equation}
	h = w\cdot \xi - \beta
\end{equation}
is greater than zero, and $c_2$ otherwise.
Since this yields exactly the same classification result as the prototype-based method described in the previous paragraph for $n=2$, this procedure can equivalently be interpreted as constructing a linear discrimination hyperplane between a pair of novel object categories and evaluating its generalizability to test samples.

Hereafter, we refer to the case of $n=2$ as \textit{pairwise few-shot learning}, and the case of $n>2$ \textit{multi-class few-shot learning}. In pair-wise few-shot learning evaluation, a confusion matrix is generated by iteratively conducting evaluations for all possible pairs of novel categories, whereas is multi-class few-shot learning, a confusion matrix is constructed from a single evaluation involving all novel categories as candidate object classes. The results of the pairwise few-shot learning evaluation are presented in Section~\ref{sec:few_shot_results}. The multi-class few-shot learning evaluation is used to compare the error patterns of DCNNs with those of human participants, and the corresponding results are shown in Section~\ref{sec:matrix_based_results}.

\subsection{Dataset\label{sec:dataset}}
\subsubsection{Image dataset: CIFAR-100\label{sec:dataset_cifar100}}
In the pre-training phase and evaluation of pairwise few-shot learning performance, we utilized the CIFAR-100 dataset \cite{krizhevsky2009learning}. This dataset consists of 60000 colored images of objects each with a resolution of 32 x 32 pixels. In this dataset, each image has two different labels to annotate which category the object in the image belongs to, namely \textit{fine category} and \textit{coarse category}. The number of fine categories defined in the dataset is 10, and each fine category belongs to one of 20 coarse categories. The number of included fine categories in each coarse category is 5. For instance, the coarse category \textit{large carnivores} include \textit{bear, leopard, tiger, wolf}, and \textit{lion}.
In both the pre-training phase and the evaluation of pairwise few-shot learning performance, we used the CIFAR-100 dataset \cite{krizhevsky2009learning}. This dataset contains 60,000 color images of objects, each with a resolution of 32x32 pixels. Each image is annotated with two levels of category labels: a \textit{fine category} and a \textit{coarse category}. There are 100 fine categories in total, grouped into 20 coarse categories, with each coarse category containing 5 fine categories. For example, the coarse category \textit{large carnivores} includes the fine categories \textit{bear}, \textit{leopard}, \textit{tiger}, \textit{wolf}, and \textit{lion}.

\subsubsection{Human visual classification task dataset: CIFAR-10H\label{sec:cifar10h}}
The dataset used to evaluate the similarity of error patterns between the DCNNs and human participants is CIFAR-10H \cite{Battleday2020-xf}. This dataset was collected in a behavioral experiment in which 2,750 human participants classified images from the well-known CIFAR-10 \cite{krizhevsky2009learning} dataset into 10 object categories. Participants were instructed to select the object category for each image as quickly as possible after its presentation. Although humans are expected to perform this task more accurately than machines, some misclassification errors were inevitably observed.

The dataset provides the results of the behavioral experiment, with the number of human participants classifying each image into each of the 10 categories. By averaging the histograms of these classifications within each ground-truth object category, we generate a misclassification pattern histogram for that category. These misclassification histograms were then arranged to form a confusion matrix representing behavior of human participants. Assuming that this confusion matrix reflects the similarity relationship between object categories in human recognition, we later compare it with the error pattern matrices obtained from the DCNNs to assess the correspondence between the DCNNs' internal representations and those of humans (see Section~\ref{sec:matrix_based_results} for the results).

\subsubsection{Categories in the datasets}
To evaluate few-shot novel category discrimination, we first define ``known categories'' as those present during the pre-training phase, and ``novel categories'' as those absent in it. For the CIFAR-100 dataset, we randomly divided the 20 coarse categories into two subsets (10 coarse categories for each) and then assigned the corresponding fine categories based on this division. Pre-training (either contrastive or supervised) was conducted using only input images belonging to the known categories. This ensures that none of the novel categories used during the evaluation of few-shot discrimination were encountered by the network during pre-training.

In addition, we conducted a separate few-shot discrimination test to generate error pattern matrices for comparison with human semantics and recognition. For this purpose, CIFAR-100 categories could not be used, as the human confusion matrices were constructed based on the category definitions of the CIFAR-10 dataset. Importantly, the category and image sets of CIFAR-10 are completely disjoint from those of CIFAR-100. This guarantees that the CIFAR-10 images also represent novel categories from the perspective of the pre-trained network.

\subsection{Comparison between the models and human semantics and recognition\label{sec:comparison_methods}}
The primary objective of this study was to evaluate the similarity between the inter-categorical structure of internal representations in DCNNs and that of human semantics and recognition. To this end, we employed two complementary methods of evaluation. The following subsections describe these methods in detail.

\subsubsection{Clustering-based evaluation: comparison with human-defined categorical aggregation}
This approach evaluates the similarity between the inter-category structure of internal representations in the networks and the semantic relationships among object categories defined by humans. As introduced in Section~\ref{sec:dataset_cifar100}, the CIFAR-100 dataset provides both fine-grained and coarse-grained category labels for each image. While fine categories are used in training and few-shot evaluation, coarse categories define broader semantic groupings of fine categories as annotated by humans. In this evaluation, we treat these coarse categories as reflecting human-defined semantic similarity between fine categories. Specifically, we assume that fine categories belonging to the same coarse category are semantically more similar to each other than to those belonging to different coarse categories.

After performing pairwise few-shot learning evaluation using the CIFAR-100 dataset, we conducted hierarchical clustering on the resulting error pattern matrix. Because an error pattern matrix reflects the similarity between internal representations of different fine categories, each resulting cluster can be interpreted as a group of fine categories that are represented more closely to each other than to those in other clusters.

Based on the resulting dendrogram, we determined a distance threshold at which the number of clusters becomes exactly 10, matching the number of coarse categories. For this set of clusters, we first performed a qualitative evaluation by examining the histogram of the number of coarse categories represented within each cluster. In addition, we quantitatively assessed the degree to which fine categories within each cluster are aligned with the human-defined coarse categories.

Let $\mathcal H$ denote the set of clusters obtained at the chosen threshold of inter-cluster distance. The threshold was selected such that the number of clusters satisfies $|\mathcal H| = 10$. Given $\mathcal H$, we computed the mutual information between the clustering results and the ground-truth coarse category set $\mathcal{C}^\text{coarse}$ as follows:
\begin{equation}
    \label{eq:mutual_information}
    I[\mathcal H; \mathcal C^\text{coarse}] = \frac{1}{|\mathcal H|} \sum_i\sum_j P(C_j^\text{coarse}|H_i) - S(\mathcal C^\text{coarse}),
\end{equation}
where $H_i\in\mathcal H$ denotes a cluster. Here, $S(\mathcal C^\text{coarse})$ is the entropy of the coarse category distribution over all fine categories. The conditional probability $P(C_j^\text{coarse}|H_i)$ is computed as a normalized histogram of coarse categories associated with the fine categories in cluster $H_i$.

\subsubsection{Matrix similarity evaluation: comparison to human confusion matrix on CIFAR-10}
In this evaluation, we compare the classification performance of the networks with that of human participants, based on the CIFAR-10H dataset \cite{Battleday2020-xf} (see Section~\ref{sec:cifar10h}). This dataset contains results from a behavioral experiment in which human participants were asked to classify images from the CIFAER-10 dataset into 10 object categories. Using these experimental data, we constructed a pseudo-confusion matrix that reflects the ``average human perception'' for this categorization task.

To evaluate the similarity between the internal representations of the models and human recognition, we computed Spearman's rank correlation between the human pseudo-confusion matrix and the error pattern matrices produced by the networks in the multi-class few-shot learning. This analysis quantifies the correspondence between inter-category similarity as perceived by humans and the representational similarity of categories in the neural networks.

\section{Results}
\subsection{Performance on few-shot novel category discrimination\label{sec:few_shot_results}}
\begin{figure}
    \centering
    \includegraphics[width=\linewidth]{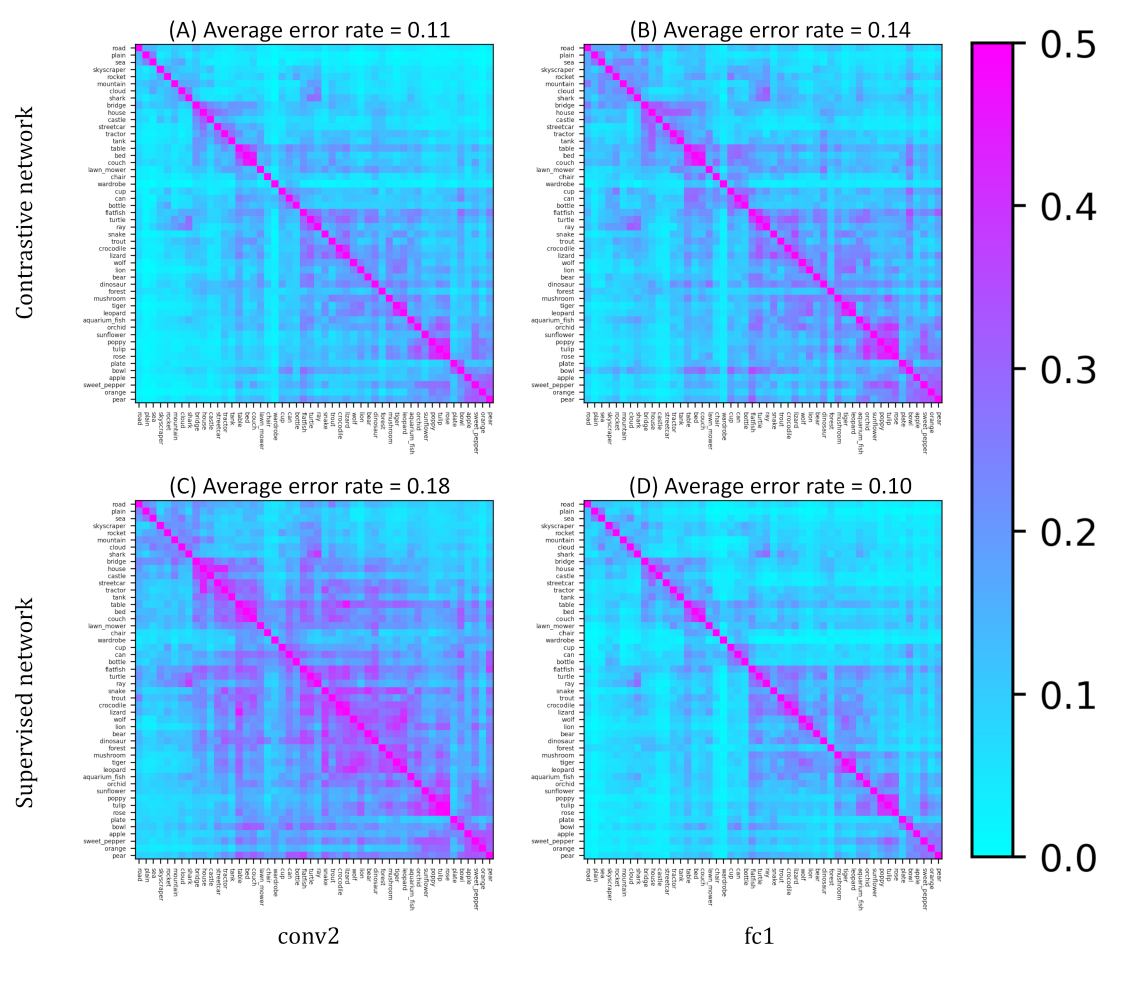}
    \caption{Error pattern matrices of the networks in representative layers. The upper panels (A and B) show the results on the network trained by contrastive learning, and the lower panels (C and D) are the results of the supervised baseline. The panels on the left (A and C) and right (B and D) show the results from the shallower conv2 layer and deeper fc1 layer, respectively. While the supervised baseline model showed slightly lower accuracy in the shallower layer, the DCNN trained by contrastive learning exhibited accurate discriminations of novel object categories in both layers.}
    \label{fig:few_shot_accuracies_matrix}
\end{figure}
\begin{figure}
    \centering
    \includegraphics[width=\linewidth]{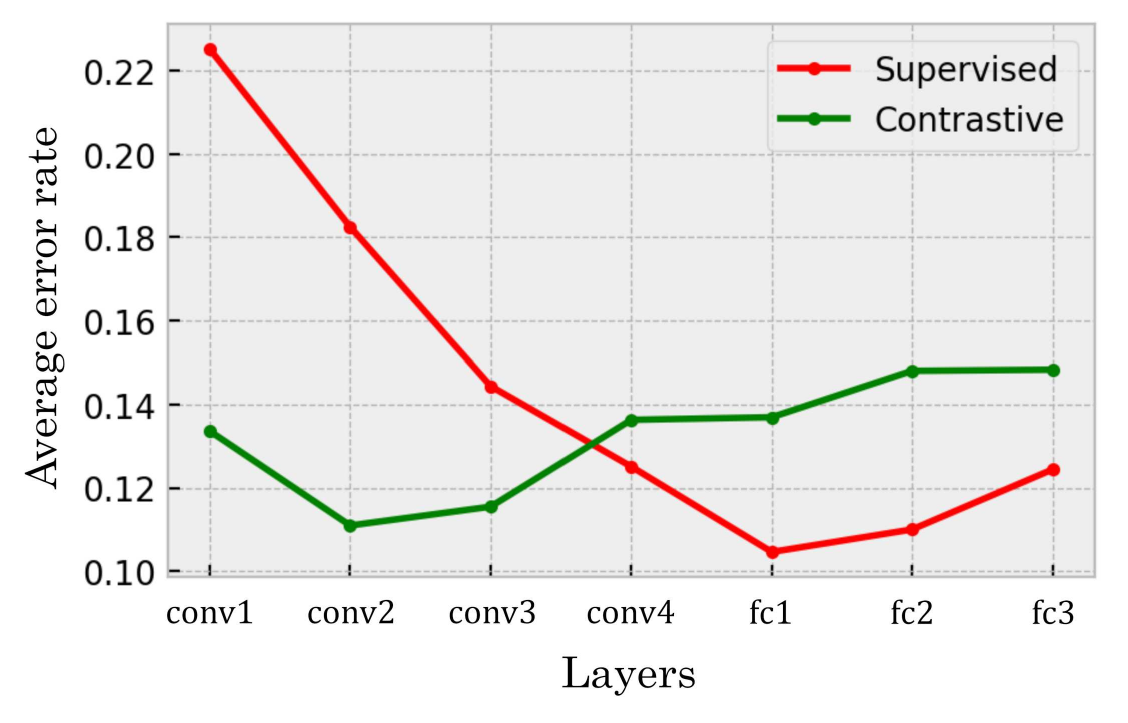}
    \caption{Mean error rate of the networks in different layers on the pairwise few-shot learning task. The green line indicates the error rates in the DCNN trained with self-supervised contrastive learning. In contrast, the red line corresponds to error rates in the supervised baseline. While the supervised baseline model provided higher accuracy in the deeper layers, the contrastive model also exhibited high accuracy along the hierarchy of the network.}
    \label{fig:few_shot_accuracies_plot}
\end{figure}

First, we evaluated the performance of a DCNN trained with self-supervised learning (SimCLR) on the task of pairwise few-shot discrimination of novel object categories, and constructed error pattern matrices from the results. As a baseline for comparison, we also evaluated a DCNN trained with supervised learning. In the pre-training phase, both networks were trained on image samples from 50 known object categories out of the 100 pre-defined categories in the CIFAR-100 dataset. After pre-training, we assessed few-shot discrimination performance using image samples from the remaining 50 novel categories.

We first present the error pattern matrices for the few-shot discrimination task involving the novel object categories, computed from the internal representations at two layers: convolutional layer 2 (conv2) and fully connected layer 1 (fc1) (Fig.~\ref{fig:few_shot_accuracies_matrix}). These layers were selected as representative examples: conv2 is a relatively shallow layer situated mid-way through convolutional processing, while fc1 is the first layer following the transition from convolutional to fully connected processing. Each element in the matrix indicates the error rate for discriminating a pair of novel categories, averaged over multiple trials using different few-shot samples. Bluish elements correspond to category pairs with error rates below approximately $15\%$ (i.e., accuracy above $85\%$), whereas reddish elements indicate near-chance-level performance ($\sim 50\%$). In the DCNN trained with self-supervised contrastive learning, most category pairs were discriminated with accuracy exceeding $80\%$. The average accuracies in conv2 and fc1 were approximately $89\%$ and $86\%$, respectively. Although there was a slight difference in performance between the two layers, no drastic degradation or improvement was observed. In contrast, the baseline DCNN trained with supervised learning exhibited a higher accuracy at the deeper fc1 layer (approximately $90\%$) compared to the shallower conv2 layer. For more detailed results across all layers, see Supplementary Fig. S2.

For a more detailed comparison between the self-supervised and supervised models in terms of layer-wise performance differences, we present the average error rates computed across all layers in each model in Fig.~\ref{fig:few_shot_accuracies_plot}. Each point in the graph represents the mean value of an error pattern matrix, excluding the diagonal elements, calculated from the representations at a specific layer of the network. In the self-supervised model, the accuracy of few-shot novel category discrimination did not show strong dependence on layer depth; performance remained relatively stable across the hierarchy. In contrast, the supervised model exhibited a clear trend: the average error rate was higher in the shallower layers and gradually decreased in the deeper layers. Although the self-supervised model outperformed the supervised model in the shallower layers, both models achieved similar levels of accuracy at their respective best-performing layers. We also evaluated a DCNN trained using SimSiam, another contrastive learning algorithm. The results were qualitatively similar to those of SimCLR, although the overall accuracy of the SimSiam model was lower (see Supplementary Fig. S1,A).

\subsection{Correspondence between unsupervised clustering of the DNN's representations and human semantic object categories\label{sec:clustering_based_results}}
\begin{figure}
    \centering
    \includegraphics[width=\linewidth]{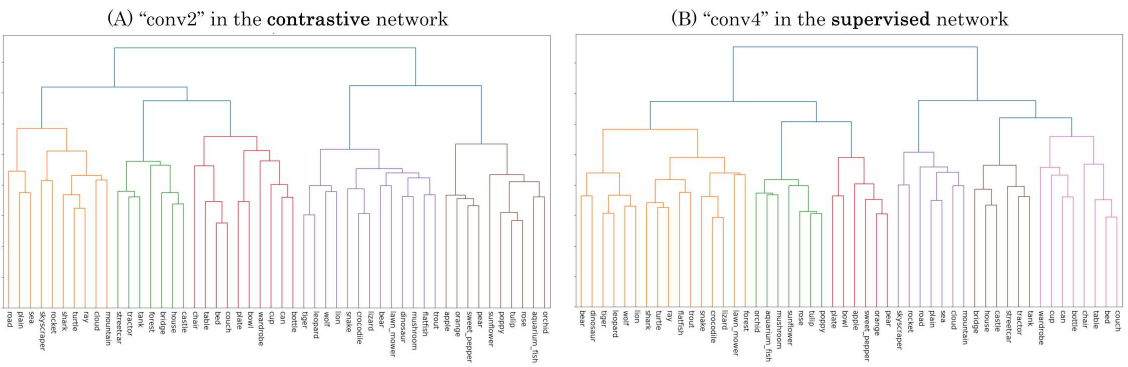}
    \caption{Dendrograms resulting from hierarchical clustering on error pattern matrices of pairwise few-shot learning. (A) The supervised network. (B) The contrastive network.}
    \label{fig:dendrograms}
\end{figure}
\begin{figure}
    \centering
    \includegraphics[width=\linewidth]{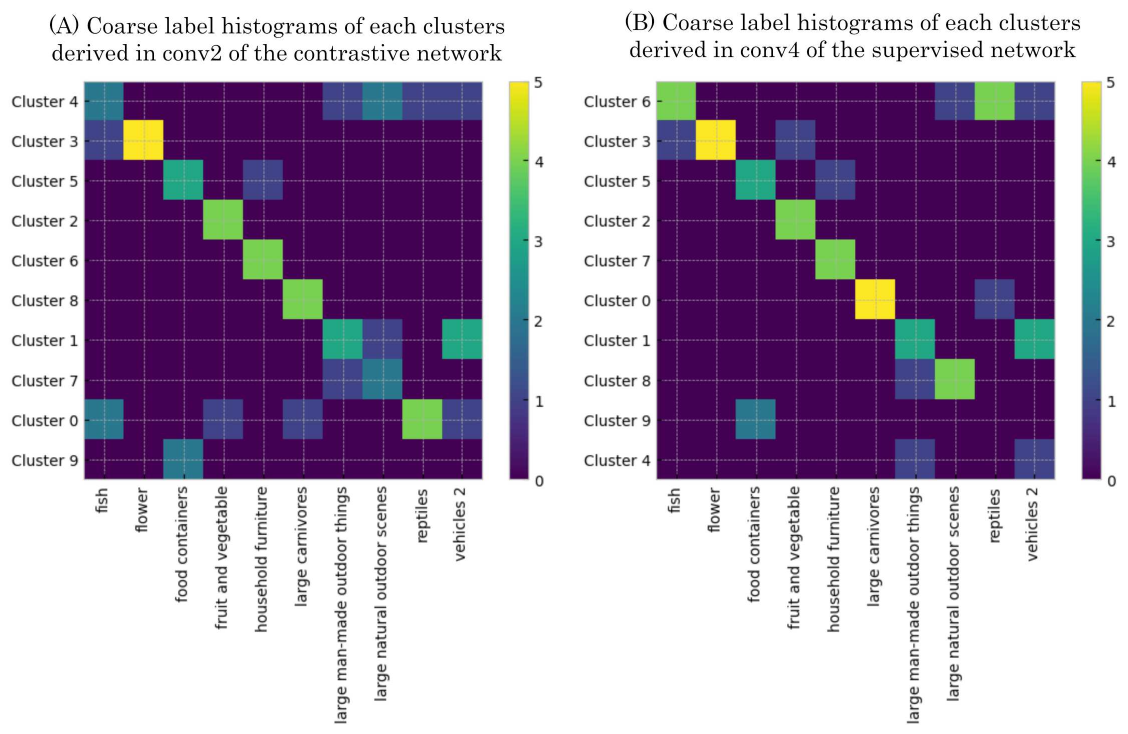}
    \caption{Derived relationships between hierarchical clusters on error pattern matrices of the networks and coarse labels. Rows correspond to clusters, wherein each component is the number of fine categories belonging to a cluster and to a coarse category at the same time. (A) Conv2 of the contrastive network. (B) Conv4 of the supervised network.}
    \label{fig:histograms}
\end{figure}
\begin{figure}
    \centering
    \includegraphics[width=\linewidth]{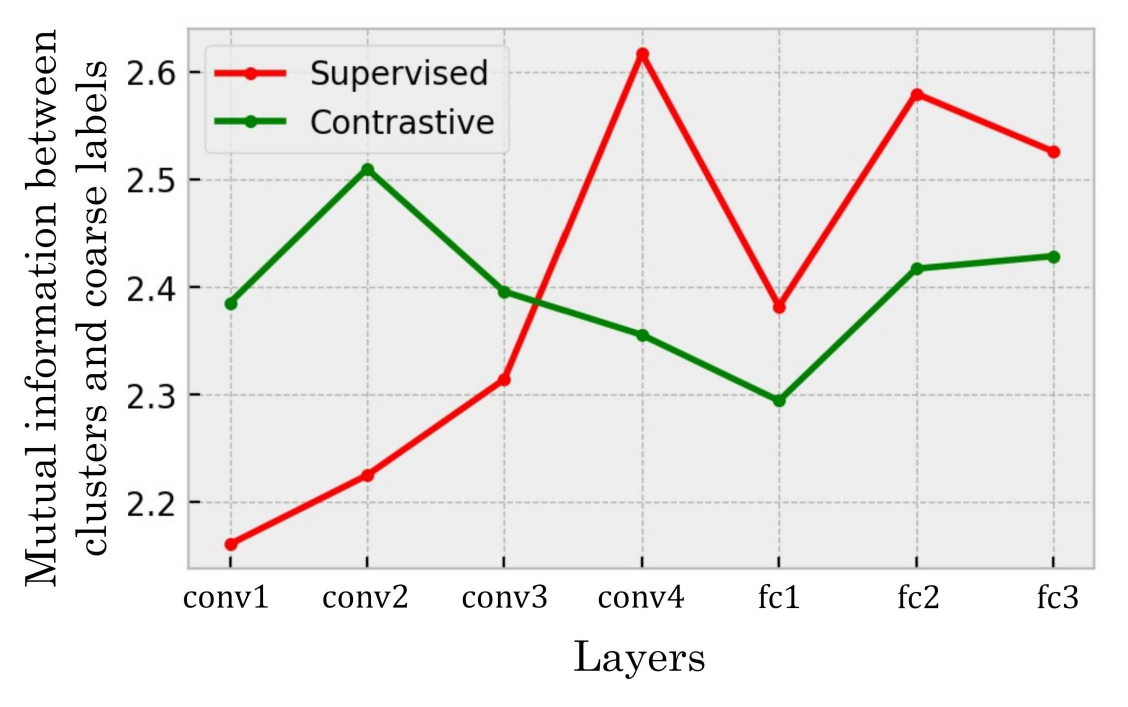}
    \caption{Mutual information between hierarchical clusters and coarse categories. The variability of the mutual information was not quite large in both of the DCNNs along their hierarchies and exhibited high values. This implies that both of the DCNNs have high similarity of resulting clusters of category representations to coarse-grained concepts defined based on human semantics.}
    \label{fig:mutual_information}
\end{figure}

Next, to examine whether the representations in the self-supervised DCNN reflect human-like semantic structures, we investigated the correspondence between the representational similarity structure of novel categories in the DCNN and the semantic relationships among categories defined by humans. To assess representational similarity in terms of few-shot learning, we interpreted the error rates shown in Fig.~\ref{fig:few_shot_accuracies_matrix} as representational distances between novel categories. To extract category groupings from these representations, we applied hierarchical clustering to the error pattern matrices (see Fig.~\ref{fig:dendrograms}).

We then evaluated how many coarse categories are represented within each resulting cluster: a cluster is considered to more strongly correspond to a certain coarse category if it contains fewer of them. Since categories within the same coarse category can be regarded as semantically similar, we assessed whether categories grouped closely in the DCNN also shared semantic similarity as defined by humans.

Fig.~\ref{fig:histograms} illustrates that the clusters of object categories derived from the DCNN representations closely matched the pre-defined coarse categories. The correspondence matrices in Fig.~\ref{fig:histograms} were computed from the layer in each network that showed the strongest alignment with human semantics. In Figs.~\ref{fig:histograms}A and B, each row represents a cluster of categories, and each column represents a human-defined coarse category. Rows were sorted so that high-value entries are aligned along the diagonal, maximizing the total sum of diagonal values. Each matrix entry indicates the number of fine-grained object categories in a cluster that belong to each coarse category. For more detailed layer-wise results, see Fig. S3.

In the self-supervised model (Fig.~\ref{fig:histograms}A, conv2), diagonal elements had consistently higher values than off-diagonal ones, suggesting that most clusters were highly aligned with specific coarse categories. Even in clusters with weaker correspondence to coarse categories, the included categories were semantically coherent--for instance, clusters rarely mixed coarse categories like ``artifacts'' and ``natural objects''. These results indicate that the self-supervised DCNN grouped novel categories in ways that are consistent with human semantic similarity. The supervised model showed a broadly similar pattern, except that a clear cluster corresponding to the ``reptiles'' category, which was observed in the self-supervised model, was missing. 

To quantify the consistency between representational clusters and human semantics, we computed the mutual information between clusters and coarse categories in each layer (see Section~\ref{sec:comparison_methods}). Fig.~\ref{fig:mutual_information} shows the layer-wise mutual information for both networks. In the self-supervised model, high mutual information values were observed throughout the hierarchy, while in the supervised model, it increased with layer depth, consistent with the trend observed in Fig.~\ref{fig:few_shot_accuracies_plot}. Although there were some differences in layer-wise patterns, the overall mutual information was comparable between the two models. Similar results were also observed in the SimSiam-trained DCNN (see Supplementary Fig. S1,B).

\subsection{Similarity between the error patterns of classification in the DCNNs and human behavioral data\label{sec:matrix_based_results}}
\begin{figure}
    \centering
    \includegraphics[width=\linewidth]{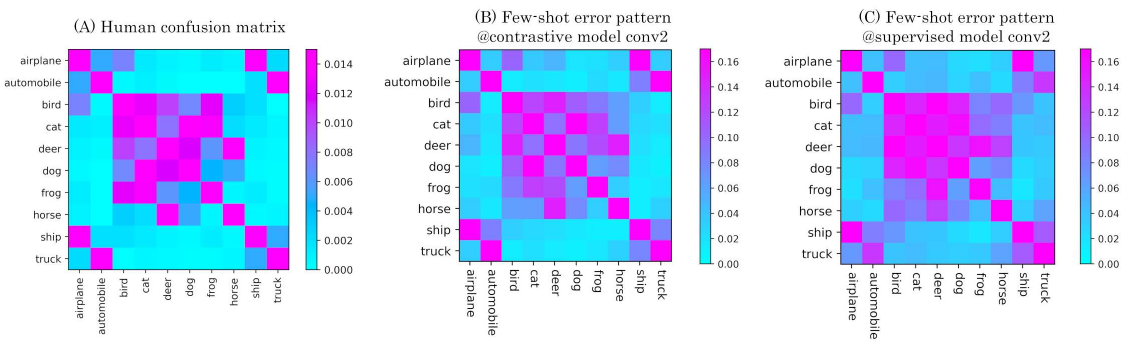}
    \caption{Confusion matrices on 10-class object categorization of the CIFAR-10 dataset. (A) Average performance of 2750 human participants. (B and C) Confusion matrices of the networks in a shallower layer, corresponding to the results of the supervised and the contrastive models, respectively. In shallower layers, we observed a higher similarity to human error patterns in the contrastive model.}
    \label{fig:cifar10h_matrix}
\end{figure}
\begin{figure}
    \centering
    \includegraphics[width=\linewidth]{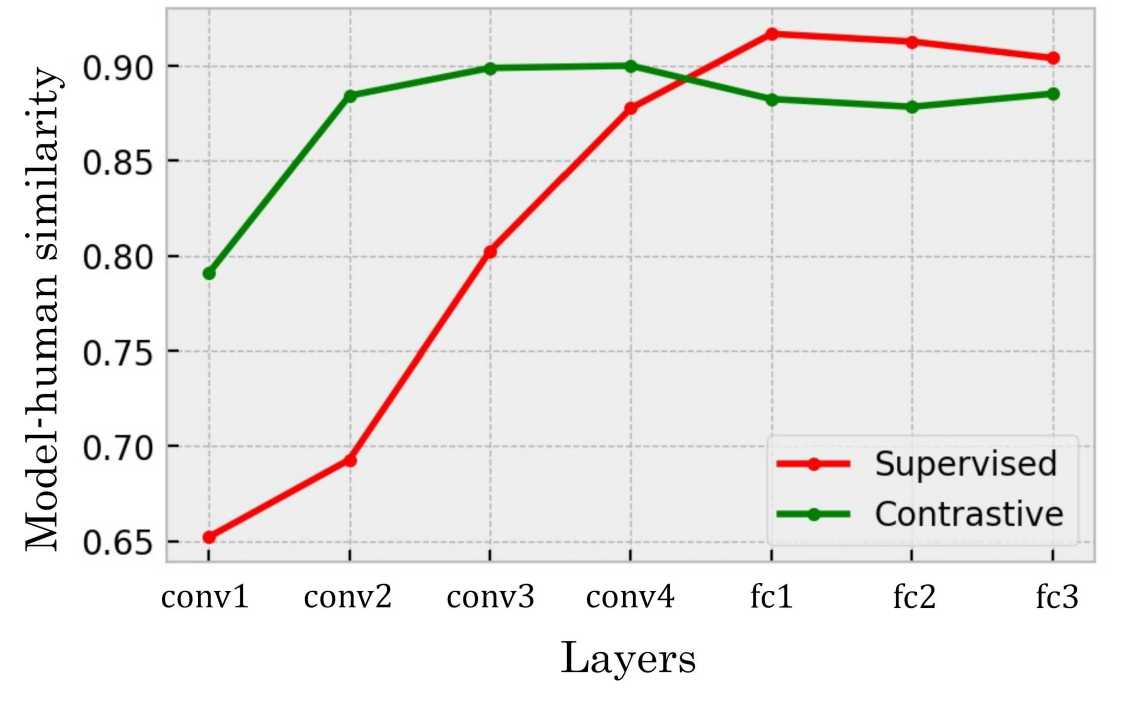}
    \caption{Spearman's rank correlations between models' confusion matrices and that of human participants computed from CIFAR-10H annotations. The contrastive DCNN exhibited high values in both shallower and deeper layers, while the supervised baseline showed lower similarity to human behavioral data in the shallower layers.}
    \label{fig:cifar10h_plot}
\end{figure}

Based on the findings from the previous subsections, we next investigated whether the similarity between object categories in DCNNs was consistent with human perception or recognition. To this end, we used the CIFAR-10H dataset \cite{Battleday2020-xf}, which contains behavioral data from human participants performing a 10-class object classification task on the CIFAR-10 dataset. We compared the confusion matrices derived from DCNNs performing multi-class few-shot learning on CIRFAR-10 images with the confusion matrix computed from human responses provided by CIFAR-10H. Similarity was quantified using Spearman's rank correlation coefficient. Note that the CIFAR-10 object categories do not overlap with those in CIFAR-100 used for training the DCNNs, guaranteeing that the object categories in this dataset are also novel to the networks.

Fig.~\ref{fig:cifar10h_matrix} (left) shows the confusion matrix generated from human behavioral data. As adult participants generally performed this task with high accuracy, the resulting confusion matrix exhibits relatively small error values. When comparing this to the confusion matrix of the self-supervised model (Fig.~\ref{fig:cifar10h_matrix}, middle), we observe similar global patterns. In particular, characteristic confusion structures present around the center and at the four corners of both matrices. These patterns suggest that object category pairs which are difficult for humans to perceptually discriminate are also similarly represented in the self-supervised model.

The supervised baseline model also produced a confusion matrix resembling the human pattern, but with generally higher error rates. This increase in errors obscured finer details in the confusion structure, resulting in a weaker alignment with both the human and self-supervised model matrices.

To more precisely evaluate the similarity between DCNNs and human perception, we computed the rank correlations across network layers. Fig.~\ref{fig:cifar10h_plot} shows the correlation coefficients for each layer. In the contrastive model, correlation values ranged from approximately 0.8 to 0.9 throughout the hierarchy, indicating a strong and stable alignment with humans' perceptual similarities between object categories. The DCNN trained with SimSiam also exhibited a similar trend (see Supplementary Fig. S1,C).

In contrast, the supervised model showed lower correlation in early layers, with values increasing in the deeper layers. Under this layer-dependent variation, correlation coefficients in the deeper layers were comparable to those of the self-supervised model.

\section{Discussion}

In this study, we investigated whether deep convolutional neural networks (DCNNs) trained with self-supervised learning could acquire internal representations that resemble those of human semantic understanding and perceptual recognition. To this end, we evaluated the networks' performance on few-shot learning tasks involving novel object categories. 

Our findings revealed three main results. First, internal representations learned through self-supervised contrastive learning (1) enabled accurate few-shot classification of novel object categories. Second, these representations (2) exhibited inter-categorical structures that closely mirrored human semantic organization. Third, they (3) produced error patterns in few-shot classification tasks that were similar to those observed in human object recognition.

\subsection{Internal representations of self-supervised learning}
Here, we discuss the non-trivial aspects of the internal representations obtained through self-supervised learning. Our findings that self-supervised contrastive learning can yield internal representations enabling accurate few-shot classification, and that the inter-categorical structure of these representations aligns with human semantic and perceptual recognitions are far from obvious.

This is because contrastive learning, particularly in the SimCLR framework, is designed to pull together positive pairs and push apart negative pairs, without any access to object category labels (as reflected in the objective function; Eq.~\ref{eq:info_nce_1} and \ref{eq:info_nce_2}). Therefore, there is no explicit reason why such training should result in representations that are both categorical and aligned with human semantics. In fact, the emergence of categorical structure through self-supervised learning may appear even more non-trivial than in supervised learning, where explicit category information is provided and thus encourages such structure. It is worth noting, however, that even in supervised learning, the acquisition of categorical representations for novel, unseen categories is not guaranteed or trivial \cite{Sorscher2022-dr}.

Furthermore, our comparisons between DCNNs trained via self-supervised and supervised learning revealed additional non-trivial findings. Across both few-shot learning performance and correspondence to human perception, we observed qualitative similarities (e.g., error patterns in Fig.~\ref{fig:few_shot_accuracies_matrix}, \ref{fig:cifar10h_matrix}; clustering structures in Fig.~\ref{fig:dendrograms}, \ref{fig:histograms}) as well as quantitative ones (e.g., mean error rates in Fig.~\ref{fig:few_shot_accuracies_plot}, mutual information in Fig.~\ref{fig:mutual_information}, and rank correlations with human confusion matrices in Fig.~\ref{fig:cifar10h_plot}).

Given the substantial difference in training objectives between self-supervised and supervised learning, these converging results are highly non-trivial and suggest a remarkable similarity in the internal representations learned by both approaches. While several theoretical connections between supervised and self-supervised objectives have been proposed \cite{Arora2019-cu, pmlr-v162-bao22e, Nozawa2021-rb}, there is currently no comprehensive theoretical explanation for the observed alignment in representations between models trained with these distinct objectives. Further theoretical investigation is needed to clarify why and how such similarities emerge between contrastive and supervised learning.

These insights raise the possibility that aspects of human semantic understanding may emerge in the absence of explicit external supervision. This idea naturally leads to the discussion in Subsection~\ref{sec:language_discussion}, where we explore the implications of these findings for language acquisition and development.

\subsection{Formation of semantics in humans\label{sec:language_discussion}}

Although we focused on the visual processing in the DCNNs and their internal representations of images and objects acquired through different learning mechanisms, the results can be interpreted in relation to the structure of human language. Specifically, our pairwise few-shot learning evaluation in Section~\ref{sec:few_shot_results} was conducted using category labels from the CIFAR-100 dataset, which are based on English vocabulary. From this perspective, the experiment can be interpreted as a test of whether visual object categories referred to by different English terms are linearly separable within the network's internal representation. Our results demonstrated that the self-supervised DCNN performed few-shot learning successfully, implying that the internal representations contain categorical structures aligned with human linguistic categorization.

Additionally, the comparison of the clusters in the networks' internal representations with coarse-grained category labels (Sec.~\ref{sec:clustering_based_results}) also provided an implication for understanding the human languages. These coarse categories used in the investigation, also derived from the English language, were found to correspond well with the clusters formed in the DCNN's internal representations. This again suggests a correspondence between the structure of language-based categories and the internal representations formed through self-supervised learning.

Based on these findings, we speculate that the categorical structure of language might, at least in part, emerge from the separability of object representations in the brain--representations that may be shaped through self-supervised learning. While the precise structure of these representations can vary depending on the learning environment and input statistics, it is plausible that self-supervised learning yields common, structured representations across individuals, which in turn inform the emergence of linguistic categories. 

Conversely, the reverse direction of influence, where language shapes perceptual recognition and even neural representation, has also been widely discussed. A prominent example is the Sapir-Whorf hypothesis \cite{Whorf2012-gm, Brutyan1969-lq, Kay1984-ks}, which posits that the structure of language can shape and even constrain cognitive perception. For instance, the conflation of ``butterflies'' and ``moths'' under the single French term \textit{papillon} may, under this hypothesis, blur perceptual distinctions for native French speakers. Empirical studies have shown that native speakers of different languages may differ in their perception of objects, time, color, and other aspects of experience that are linguistically encoded \cite{Boroditsky2001-xr, Lupyan2020-de}. Although we did not directly address this reverse effect in our study, it remains an important direction for understanding how language influences the development of neural representations.

Taking both directions into account, it seems reasonable to hypothesize that neural representations are initially formed through self-supervised learning during early development such as infancy, and subsequently fine-tuned by language-based supervision. Most computational studies to date have focused on the outcome of a single learning rule. To better understand brain-like learning mechanisms, future research should consider how the interplay between self-supervised learning and supervised fine-tuning models the developmental progression of neural representations from infancy to adulthood.

\subsection{Towards a more biologically plausible learning mechanism\label{sec:prediction_discussion}}
Here, we discuss the implications of our findings for understanding the formation of categorical representations in biological neural systems. The central result of this study is that a DCNN trained with self-supervised contrastive learning can develop internal representations of visual objects closely resembling human perceptual recognition and semantic organization. If a similar mechanism operates in biological brains, abstract categorical representations might be naturally formed prior to language-based learning. Below, we first address how contrastive learning might be biologically implemented through prediction-based learning mechanisms and then discuss how biologically plausible visual input augmentations naturally arise from such mechanisms.

A plausible implementation of contrastive learning in biological brains would be prediction-based learning, a central component in many theoretical neural processing frameworks \cite{Rao1999-uv, Friston2010-ya}. Prior studies have formalized contrastive learning using predictive paradigms by defining temporally proximal events as positive pairs and distant events as negative pairs \cite{Van_den_Oord2018-sy, Lowe2019-hz, noauthor_undated-hr}. Unlike SimCLR, which explicitly contrasts positive and negative pairs within the same batch, prediction-based learning naturally distinguishes positive and negative pairs through temporal proximity without explicit negative sampling or specific architectural constraints. Despite these differences, both SimCLR and biologically plausible prediction-based learning fundamentally share the principle of forming structured representations by comparing related and unrelated experiences, highlighting the biological relevance of the computational principles demonstrated by SimCLR in our study.

Furthermore, if we regard prediction-based learning as a plausible candidate, the visual input augmentations integral to contrastive learning such as image rotations or random cropping can naturally occur through bodily movements and sensorimotor interactions, in addition to natural temporal changes in the input from the external environment. Although the artificial augmentations used in this study include those that might not exactly correspond to natural conditions (color distortion, grayscaling, or random blurring), the remaining transformations commonly occur in biological contexts through movements such as head rotations, locomotion, and saccadic eye movements. For instance, neck rotations cause corresponding rotations in retinal images, and moving closer to an object results in a visual effect analogous to cropping. Thus, sensorimotor experiences encountered in early development inherently provide the biological basis for visual augmentations that parallel those used in computational contrastive learning.

Taken together, our finding that abstract, human-like categorical representations can emerge from self-supervised contrastive learning provides a promising basis for understanding how such representations may form in the human brain without explicit supervision. If biologically plausible learning mechanisms--such as prediction-based learning shaped by natural sensorimotor experience--can approximate contrastive learning, as discussed above, then our results suggest that conceptual representations could arise through self-supervised processes alone. Rather than claiming that the brain implements contrastive learning per se, our study identifies a representational target and computational principle that future biologically grounded models can aim to approximate. This offers a concrete step toward linking the unsupervised emergence of conceptual structure in artificial systems to that in biological neural systems.

\section*{Acknowledgments}
A.K. is supported by JSPS KAKENHI Grant Number 24KJ0798. M.O. is supported by JST Moonshot R\&D Grant Number JPMJMS2012, and JSPS KAKENHI Grant Numbers 20H05712 and 23H04834.

\appendix

\section{SimSiam contrastive learning}
In this section, we introduce SimSiam (Chen and He, 2020) as another example of self-supervised contrastive learning algorithm. By conducting the same series of experiments using another DCNN trained by SimSiam learning algorithm, we can discuss generality of the results reviewed in the main content among different self-supervised contrastive learning algorithms.

The DCNN utilized for SimSiam takes the same encoder architecture as the SimCLR model and the supervised baseline in the main content. The netwrok $f$ consists of a ResNet-18 encoder $g$ and additional NLP (fully connected layers) modules $\operatorname{proj}$.

A characteristic of SimSiam that is different from SimCLR is the loss function and the training procedure. While the network takes a positive sample $x^+$ and negative samples $\{x^-_k\}_k$ for each anchor input $\tilde x$ in SimCLR, SimSiam only takes a positive sample for each anchor. Let us take an anchor input $\tilde x$ and its positive sample $x^+$. For one input, the network computes the representation as the output of $g$, while the representation is computed as the output of $\operatorname{proj}$ for the other. The objective function is the negative cosine similarity between the two representations:
\begin{equation}
    l_\text{siam}(x, x^+) = -\left\langle\frac{f(x)}{\|f(x)\|_2}, \frac{g(x^+)}{\|g(x^+)\|_2}\right\rangle.
\end{equation}
While this metric is asymmetrical with respect to $x$ and $x^+$, the qualitative role of them can be considered symmetrical: if the positive sample $x^+$ is treated as an anchor sample, then the anchor sample $x$ can also function as a positive sample. Hence, the full expression of the loss function is the average symmetrized negative cosine similarity:
\begin{equation}
    \mathcal{L}_\text{siam} = -\mathbb{E}_{x\sim\rho_\text{im}, a_1\sim\rho_\text{aug}, a_2\sim\rho_\text{aug}}\left[\frac{l_\text{siam}(a_1(x), a_2(x)) + l_\text{siam}(a_2(x), a_1(x))}{2}\right].
\end{equation}

The optimization algorithm for SimSiam is also different from SimCLR. The algorithm is based on error back-propagation. However, the gradient of the loss function is computed only with respect to the first argument of $l_\text{siam}$. The gradient used for updates of the synaptic parameters in the DCNNs is then
\begin{equation}
    \operatorname{grad}(l_\text{siam}(x_1, x_2)) = \frac{\partial }{\partial x_1}l_\text{siam}(x_1, x_2).
\end{equation}

As mentioned above, SimSiam does not require negative samples for training. This feature reduces the computational complexity of the training procedure. Furthermore, this negative sampling-free learning framework is considered more plausible in biological systems. While SimCLR requires the memory of representations of large amount of negative samples, for SimSiam, the network does not need to store such information and only compares representations at different layers. Although the gradient stopping is still implausible, we assume the relationship of contrastive learning to prediction-based learning mechanisms discussed in the main content also holds for SimSiam.

\section{Evaluations of internal representations of SimSiam model}

\begin{figure}[t]
    \centering
    \includegraphics[width=\linewidth]{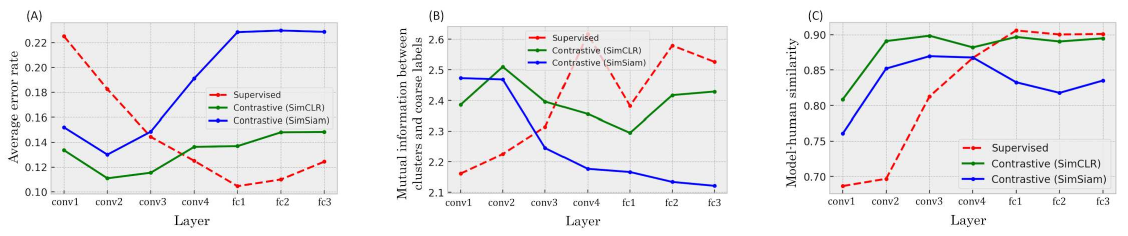}
    \caption{Evaluations of internal representations of the SimSiam model. (A) The layer-wise average error rates of pairwise few-shot learning. All the three subfigures are based on Fig \ref{fig:few_shot_accuracies_plot}, \ref{fig:mutual_information}, and \ref{fig:cifar10h_plot} in the main content. An additional result from the DCNN trained by SimSiam (the blue solid line) is added to each of the subfigures. (B) The layer-wise mutual information between hierarchical clusters of the internal representations and human-defined coarse graining of novel object categories. (C) The layer-wise rank correlation between confusion matrices of multi-class few-shot learning and of human participants.}
    \label{fig:results_with_siam}
\end{figure}

\begin{figure}[t]
    \centering
    \includegraphics[width=\linewidth]{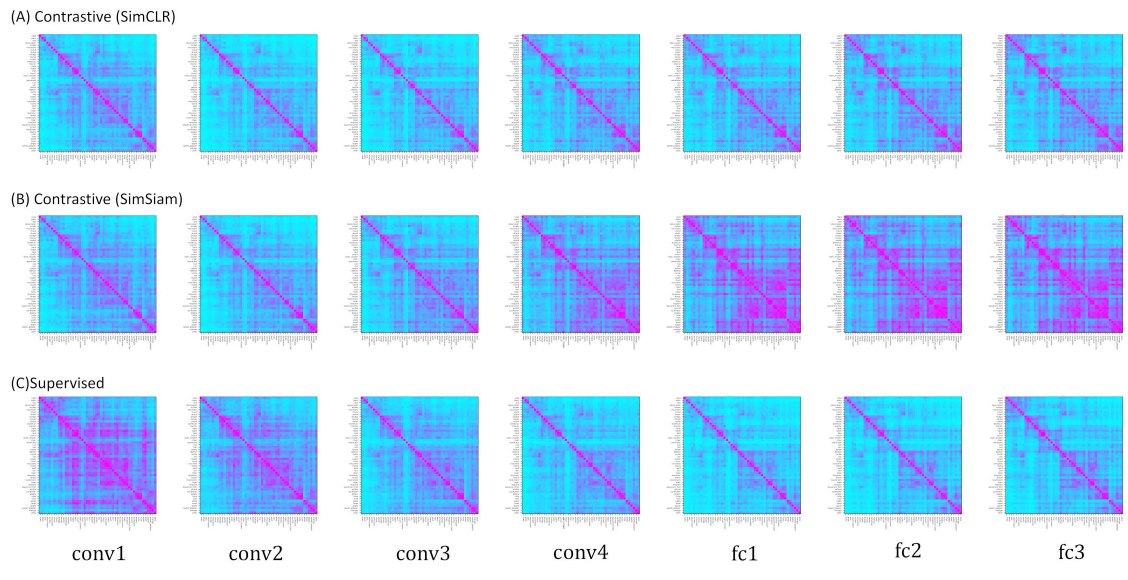}
    \caption{Error pattern matrices of pair-wise few-shot learning of internal representations from all layers of the (A) SimCLR model, (B) SimSiam model, and (C) supervised baseline. Fig \ref{fig:results_with_siam}(A) is obtained by averaging these matrices except for the diagonal elements. While the values vary between the models, the overall error pattern structures were qualitatively similar.}
    \label{fig:all_layer_fewshot}
\end{figure}

\begin{figure}[t]
    \centering
    \includegraphics[width=\linewidth]{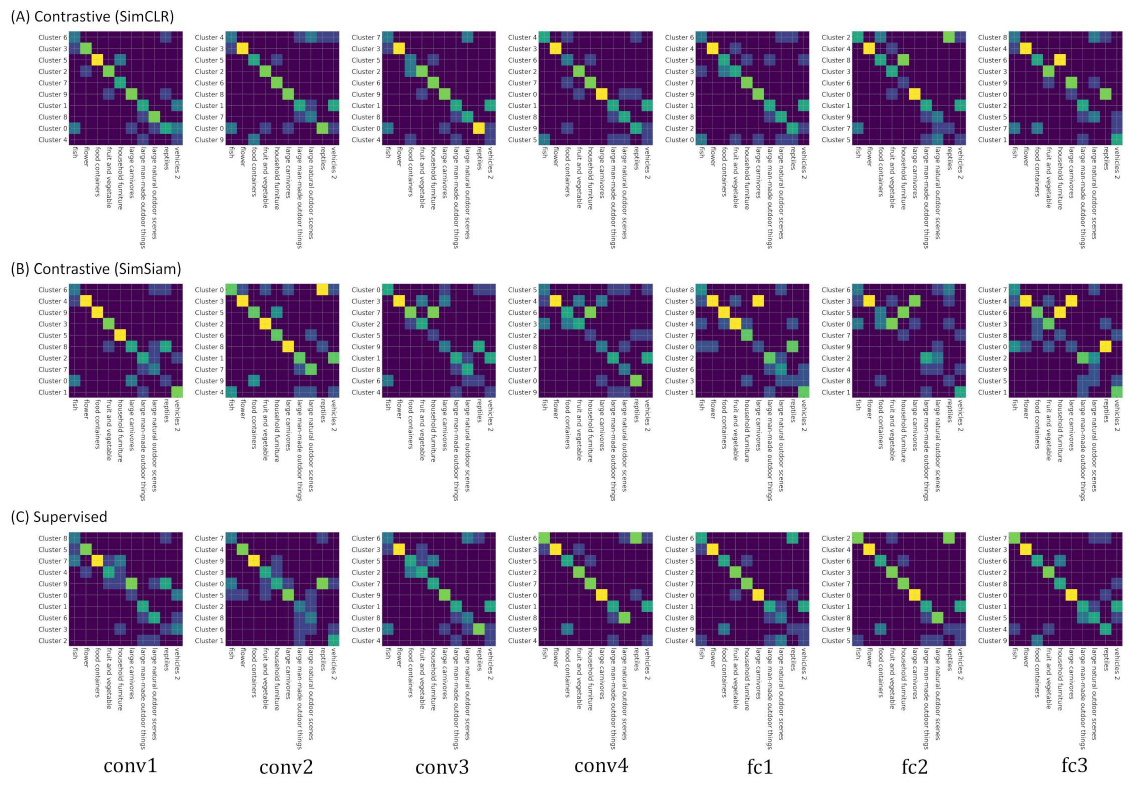}
    \caption{Correspondence of human semantic coarse categories and the clusters of internal representations from all layers of the (A) SimCLR model, (B) SimSiam model, and (C) supervised baseline. Shared tendency between the two models trained by contrastive learning algorithms is that the correspondences between the coarse object categories defined based on human semantics and the hierarchical clusters collapse as it proceeds to the deeper layers, while the supervised baseline shows the opposite tendency. Fig \ref{fig:results_with_siam}(B) can be obtained by calculating mutual information between the sets of coarse categories and hierarchical clusters from these matrices.}
    \label{fig:all_layer_histogram}
\end{figure}

Here we show the evaluations of a DCNN trained by SimSiam for few-shot learning accuracy (Fig \ref{fig:results_with_siam}A), correspondences of representations to human semantics (Fig \ref{fig:results_with_siam}B), and correspondences of representations to recognition (Fig \ref{fig:results_with_siam}C). The three panels (A), (B), and (C) in Fig \ref{fig:results_with_siam} respectively correspond to Figs 5, 8, and 10 in the main text. We show the results from the SimSiam model as the blue solid line and also show the results from the supervised model and the SimCLR model as the red dotted line and the green solid line for comparison. 

The measured accuracies and correspondences to human semantics / recognition were close to that of the DCNN trained by SimCLR in the shallower layers, while they exhibited lower values in the deeper layers (Figs. \ref{fig:results_with_siam}(A)(B)(C)). These differences between the two DCNNs trained by SimSiam and SimCLR, especially in the deeper layers, were probably caused by the different learning algorithms. Even with the SimSiam learning method, if the size of the training dataset is much larger than that used in this study, the performance of SimSiam would change and become closer to that of SimCLR, as shown in the previous study \cite{Chen2020-zq}. We leave this issue as a future work.

\bibliographystyle{unsrt}  
\bibliography{refs}

\begin{thebibliography}{10}

\bibitem{Arora2019-cu}
Sanjeev Arora, Hrishikesh Khandeparkar, Mikhail Khodak, Orestis Plevrakis, and Nikunj Saunshi.
\newblock A theoretical analysis of contrastive unsupervised representation learning.
\newblock {\em arXiv}, 1902.09229, February 2019.

\bibitem{Medina2020-iy}
Carlos Medina, Arnout Devos, and Matthias Grossglauser.
\newblock {Self-Supervised} prototypical transfer learning for {Few-Shot} classification.
\newblock {\em arXiv}, 2006.11325, June 2020.

\bibitem{Chen2020-zq}
Xinlei Chen and Kaiming He.
\newblock Exploring simple siamese representation learning.
\newblock {\em arXiv}, 2011.10566, November 2020.

\bibitem{Newell2020-ff}
Alejandro Newell and Jia Deng.
\newblock How useful is self-supervised pretraining for visual tasks?
\newblock In {\em 2020 {IEEE/CVF} Conference on Computer Vision and Pattern Recognition ({CVPR})}. IEEE, June 2020.

\bibitem{Ericsson2021-oq}
Linus Ericsson, Henry Gouk, Chen~Change Loy, and Timothy~M Hospedales.
\newblock {Self-Supervised} representation learning: Introduction, advances and challenges.
\newblock {\em arXiv}, 2110.09327, October 2021.

\bibitem{Nozawa2021-rb}
Kento Nozawa and Issei Sato.
\newblock Understanding negative samples in instance discriminative self-supervised representation learning.
\newblock {\em arXiv}, 2102.06866, February 2021.

\bibitem{Shi2021-qi}
Pengxiang Shi, Wenwen Ye, and Zheng Qin.
\newblock {Self-Supervised} pre-training for time series classification.
\newblock In {\em 2021 International Joint Conference on Neural Networks ({IJCNN})}, pages 1--8. IEEE, July 2021.

\bibitem{Wang2021-ag}
Xinlong Wang, Rufeng Zhang, Chunhua Shen, Tao Kong, and Lei Li.
\newblock Dense contrastive learning for self-supervised visual pre-training.
\newblock In {\em 2021 {IEEE/CVF} Conference on Computer Vision and Pattern Recognition ({CVPR})}. IEEE, June 2021.

\bibitem{Zhu2022-tp}
Kuan Zhu, Haiyun Guo, Tianyi Yan, Yousong Zhu, Jinqiao Wang, and Ming Tang.
\newblock {PASS}: {Part-Aware} {Self-Supervised} {Pre-Training} for person {Re-Identification}.
\newblock {\em arXiv}, 2203.03931, March 2022.

\bibitem{Hu2024-vz}
Haigen Hu, Xiaoyuan Wang, Yan Zhang, Qi~Chen, and Qiu Guan.
\newblock A comprehensive survey on contrastive learning.
\newblock {\em Neurocomputing}, 610(128645):128645, December 2024.

\bibitem{Knudsen1994-yf}
E~I Knudsen.
\newblock Supervised learning in the brain.
\newblock {\em J. Neurosci.}, 14(7):3985--3997, July 1994.

\bibitem{Glaser2019-bj}
Joshua~I Glaser, Ari~S Benjamin, Roozbeh Farhoodi, and Konrad~P Kording.
\newblock The roles of supervised machine learning in systems neuroscience.
\newblock {\em Prog. Neurobiol.}, 175:126--137, April 2019.

\bibitem{Loewenstein2021-mw}
Yonatan Loewenstein, Ofri Raviv, and Merav Ahissar.
\newblock Dissecting the roles of supervised and unsupervised learning in perceptual discrimination judgments.
\newblock {\em J. Neurosci.}, 41(4):757--765, January 2021.

\bibitem{Carey1978-nc}
Susan Carey and Elsa Bartlett.
\newblock Acquiring a single new word.
\newblock {\em Papers and Reports on Child Language Development}, 15:17--29, August 1978.

\bibitem{Quinn1993-ol}
P~C Quinn, P~D Eimas, and S~L Rosenkrantz.
\newblock Evidence for representations of perceptually similar natural categories by 3-month-old and 4-month-old infants.
\newblock {\em Perception}, 22(4):463--475, 1993.

\bibitem{Behl-Chadha1996-qx}
G~Behl-Chadha.
\newblock Basic-level and superordinate-like categorical representations in early infancy.
\newblock {\em Cognition}, 60(2):105--141, August 1996.

\bibitem{Freedman2001-md}
D~J Freedman, M~Riesenhuber, T~Poggio, and E~K Miller.
\newblock Categorical representation of visual stimuli in the primate prefrontal cortex.
\newblock {\em Science}, 291(5502):312--316, January 2001.

\bibitem{Smith2002-up}
Linda~B Smith, Susan~S Jones, Barbara Landau, Lisa Gershkoff-Stowe, and Larissa Samuelson.
\newblock Object name learning provides on-the-job training for attention.
\newblock {\em Psychol. Sci.}, 13(1):13--19, January 2002.

\bibitem{Yang2016-lc}
Jiale Yang, So~Kanazawa, Masami~K Yamaguchi, and Ichiro Kuriki.
\newblock Cortical response to categorical color perception in infants investigated by near-infrared spectroscopy.
\newblock {\em Proc. Natl. Acad. Sci. U. S. A.}, 113(9):2370--2375, March 2016.

\bibitem{Lecun1998-up}
Y~Lecun, L~Bottou, Y~Bengio, and P~Haffner.
\newblock Gradient-based learning applied to document recognition.
\newblock {\em Proc. IEEE}, 86(11):2278--2324, November 1998.

\bibitem{Kriegeskorte2008-bz}
Nikolaus Kriegeskorte, Marieke Mur, and Peter Bandettini.
\newblock Representational similarity analysis - connecting the branches of systems neuroscience.
\newblock {\em Front. Syst. Neurosci.}, 2:4, November 2008.

\bibitem{Jarrett2009-kf}
Kevin Jarrett, Koray Kavukcuoglu, Marc'aurelio Ranzato, and Yann LeCun.
\newblock What is the best multi-stage architecture for object recognition?
\newblock In {\em 2009 {IEEE} 12th International Conference on Computer Vision}, pages 2146--2153, September 2009.

\bibitem{Krizhevsky2012-mo}
Alex Krizhevsky, Ilya Sutskever, and Geoffrey~E Hinton.
\newblock Imagenet classification with deep convolutional neural networks.
\newblock {\em Adv. Neural Inf. Process. Syst.}, 25, 2012.

\bibitem{Yamins2013-pu}
Daniel Yamins, Ha~Hong, Charles Cadieu, and James~J Dicarlo.
\newblock Hierarchical modular optimization of convolutional networks achieves representations similar to macaque {IT} and human ventral stream.
\newblock {\em Neural Information Processing Systems}, 2013.

\bibitem{Yamins2014-pi}
Daniel L~K Yamins, Ha~Hong, Charles~F Cadieu, Ethan~A Solomon, Darren Seibert, and James~J DiCarlo.
\newblock Performance-optimized hierarchical models predict neural responses in higher visual cortex.
\newblock {\em Proc. Natl. Acad. Sci. U. S. A.}, 111(23):8619--8624, June 2014.

\bibitem{Khaligh-Razavi2014-an}
Seyed-Mahdi Khaligh-Razavi and Nikolaus Kriegeskorte.
\newblock Deep supervised, but not unsupervised, models may explain {IT} cortical representation.
\newblock {\em PLoS Comput. Biol.}, 10(11):e1003915, November 2014.

\bibitem{Majaj2015-nf}
Najib~J Majaj, Ha~Hong, Ethan~A Solomon, and James~J DiCarlo.
\newblock Simple learned weighted sums of inferior temporal neuronal firing rates accurately predict human core object recognition performance.
\newblock {\em J. Neurosci.}, 35(39):13402--13418, September 2015.

\bibitem{Yamins2016-mu}
Daniel L~K Yamins and James~J DiCarlo.
\newblock Using goal-driven deep learning models to understand sensory cortex.
\newblock {\em Nat. Neurosci.}, 19(3):356--365, March 2016.

\bibitem{Rafegas2018-ak}
Ivet Rafegas and Maria Vanrell.
\newblock Color encoding in biologically-inspired convolutional neural networks.
\newblock {\em Vision Res.}, 151:7--17, October 2018.

\bibitem{Rajalingham2018-su}
Rishi Rajalingham, Elias~B Issa, Pouya Bashivan, Kohitij Kar, Kailyn Schmidt, and James~J DiCarlo.
\newblock {Large-Scale}, {High-Resolution} comparison of the core visual object recognition behavior of humans, monkeys, and {State-of-the-Art} deep artificial neural networks.
\newblock {\em J. Neurosci.}, 38(33):7255--7269, August 2018.

\bibitem{Hebart2020-pv}
Martin~N Hebart, Charles~Y Zheng, Francisco Pereira, and Chris~I Baker.
\newblock Revealing the multidimensional mental representations of natural objects underlying human similarity judgements.
\newblock {\em Nat. Hum. Behav.}, 4(11):1173--1185, November 2020.

\bibitem{Marques2021-kx}
Tiago Marques, Martin Schrimpf, and James~J DiCarlo.
\newblock Multi-scale hierarchical neural network models that bridge from single neurons in the primate primary visual cortex to object recognition behavior.
\newblock {\em bioRxiv}, 2021.03.01.433495, August 2021.

\bibitem{Kawakita2024-pw}
Genji Kawakita, Ariel Zeleznikow-Johnston, Naotsugu Tsuchiya, and Masafumi Oizumi.
\newblock Gromov-wasserstein unsupervised alignment reveals structural correspondences between the color similarity structures of humans and large language models.
\newblock {\em Sci. Rep.}, 14(1):15917, July 2024.

\bibitem{Bakhtiari2021-fw}
Shahab Bakhtiari, Patrick Mineault, Timothy Lillicrap, Christopher Pack, and Blake Richards.
\newblock The functional specialization of visual cortex emerges from training parallel pathways with self-supervised predictive learning.
\newblock In M~Ranzato, A~Beygelzimer, Y~Dauphin, P~S Liang, and J~Wortman Vaughan, editors, {\em Advances in Neural Information Processing Systems}, volume~34, pages 25164--25178. Curran Associates, Inc., 2021.

\bibitem{Zhuang2021-ly}
Chengxu Zhuang, Siming Yan, Aran Nayebi, Martin Schrimpf, Michael~C Frank, James~J DiCarlo, and Daniel L~K Yamins.
\newblock Unsupervised neural network models of the ventral visual stream.
\newblock {\em Proc. Natl. Acad. Sci. U. S. A.}, 118(3):e2014196118, January 2021.

\bibitem{Nayebi2021-nv}
Aran Nayebi, Nathan C~L Kong, Chengxu Zhuang, Justin~L Gardner, Anthony~M Norcia, and Daniel L~K Yamins.
\newblock Unsupervised models of mouse visual cortex.
\newblock {\em bioRxiv}, 2021.06.16.448730, June 2021.

\bibitem{Konkle2021-ox}
Talia Konkle and George~A Alvarez.
\newblock Beyond category-supervision: instance-level contrastive learning models predict human visual system responses to objects.
\newblock {\em bioRxiv}, 2021.05.28.446118, May 2021.

\bibitem{cadena2019how}
Santiago~A. Cadena, Fabian~H. Sinz, Taliah Muhammad, Emmanouil Froudarakis, Erick Cobos, Edgar~Y. Walker, Jake Reimer, Matthias Bethge, Andreas Tolias, and Alexander~S. Ecker.
\newblock How well do deep neural networks trained on object recognition characterize the mouse visual system?
\newblock In {\em Real Neurons {\&} Hidden Units: Future directions at the intersection of neuroscience and artificial intelligence @ NeurIPS 2019}, 2019.

\bibitem{Konkle2022-wo}
Talia Konkle and George~A Alvarez.
\newblock A self-supervised domain-general learning framework for human ventral stream representation.
\newblock {\em Nat. Commun.}, 13(1):491, January 2022.

\bibitem{Millet2022-kd}
Juliette Millet, Charlotte Caucheteux, Pierre Orhan, Yves Boubenec, Alexandre Gramfort, Ewan Dunbar, Christophe Pallier, and Jean-Remi King.
\newblock Toward a realistic model of speech processing in the brain with self-supervised learning.
\newblock In Alice~H Oh, Alekh Agarwal, Danielle Belgrave, and Kyunghyun Cho, editors, {\em Advances in Neural Information Processing Systems}, 2022.

\bibitem{Prince2024-yp}
Jacob~S Prince, George~A Alvarez, and Talia Konkle.
\newblock Contrastive learning explains the emergence and function of visual category-selective regions.
\newblock {\em Sci. Adv.}, 10(39):eadl1776, September 2024.

\bibitem{Sorscher2022-dr}
Ben Sorscher, Surya Ganguli, and Haim Sompolinsky.
\newblock Neural representational geometry underlies few-shot concept learning.
\newblock {\em Proc. Natl. Acad. Sci. U. S. A.}, 119(43):e2200800119, October 2022.

\bibitem{Lu2022-wx}
Yuning Lu, Liangjian Wen, Jianzhuang Liu, Yajing Liu, and Xinmei Tian.
\newblock Self-supervision can be a good few-shot learner.
\newblock {\em Computer Vision -- ECCV 2022}, pages 740--758, 2022.

\bibitem{Jaiswal2020-dy}
Ashish Jaiswal, Ashwin~Ramesh Babu, Mohammad~Zaki Zadeh, Debapriya Banerjee, and Fillia Makedon.
\newblock A survey on contrastive {Self-Supervised} learning.
\newblock {\em Technologies}, 9(1):2, December 2020.

\bibitem{Kumar2022-gj}
Pranjal Kumar, Piyush Rawat, and Siddhartha Chauhan.
\newblock Contrastive self-supervised learning: review, progress, challenges and future research directions.
\newblock {\em Int. J. Multimed. Inf. Retr.}, 11(4):461--488, December 2022.

\bibitem{Chen2020-ih}
Ting Chen, Simon Kornblith, Mohammad Norouzi, and Geoffrey Hinton.
\newblock A simple framework for contrastive learning of visual representations.
\newblock {\em arXiv}, 2002.05709, February 2020.

\bibitem{Rumelhart1986-rm}
David~E Rumelhart, Geoffrey~E Hinton, and Ronald~J Williams.
\newblock Learning representations by back-propagating errors.
\newblock {\em Nature}, 323(6088):533--536, October 1986.

\bibitem{He2015-vx}
Kaiming He, Xiangyu Zhang, Shaoqing Ren, and Jian Sun.
\newblock Deep residual learning for image recognition.
\newblock {\em arXiv}, 1512.03385, December 2015.

\bibitem{krizhevsky2009learning}
Alex Krizhevsky and Geoffrey Hinton.
\newblock Learning multiple layers of features from tiny images.
\newblock Technical Report~0, University of Toronto, Toronto, Ontario, 2009.

\bibitem{Battleday2020-xf}
Ruairidh~M Battleday, Joshua~C Peterson, and Thomas~L Griffiths.
\newblock Capturing human categorization of natural images by combining deep networks and cognitive models.
\newblock {\em Nat. Commun.}, 11(1):5418, October 2020.

\bibitem{pmlr-v162-bao22e}
Han Bao, Yoshihiro Nagano, and Kento Nozawa.
\newblock On the surrogate gap between contrastive and supervised losses.
\newblock In Kamalika Chaudhuri, Stefanie Jegelka, Le~Song, Csaba Szepesvari, Gang Niu, and Sivan Sabato, editors, {\em Proceedings of the 39th International Conference on Machine Learning}, volume 162 of {\em Proceedings of Machine Learning Research}, pages 1585--1606. PMLR, 17--23 Jul 2022.

\bibitem{Whorf2012-gm}
Benjamin~Lee Whorf.
\newblock {\em Selected Writings of Benjamin Lee Whorf}.
\newblock The MIT Press, 2012.

\bibitem{Brutyan1969-lq}
G~A Brutyan.
\newblock On the {Sapir-Whorf} hypothesis.
\newblock {\em Problemy Filosofii}, 23(1):56--66, 1969.

\bibitem{Kay1984-ks}
Paul Kay and Willett Kempton.
\newblock What is the {Sapir-Whorf} hypothesis?
\newblock {\em Am. Anthropol.}, 86(1):65--79, March 1984.

\bibitem{Boroditsky2001-xr}
Lera Boroditsky.
\newblock Does language shape thought?: Mandarin and english speakers' conceptions of time.
\newblock {\em Cognitive Psychology}, 43(1):1--22, 2001.

\bibitem{Lupyan2020-de}
Gary Lupyan, Rasha Abdel~Rahman, Lera Boroditsky, and Andy Clark.
\newblock Effects of language on visual perception.
\newblock {\em Trends Cogn. Sci.}, 24(11):930--944, November 2020.

\bibitem{Rao1999-uv}
R~P Rao and D~H Ballard.
\newblock Predictive coding in the visual cortex: a functional interpretation of some extra-classical receptive-field effects.
\newblock {\em Nat. Neurosci.}, 2(1):79--87, January 1999.

\bibitem{Friston2010-ya}
Karl Friston.
\newblock The free-energy principle: a unified brain theory?
\newblock {\em Nat. Rev. Neurosci.}, 11(2):127--138, February 2010.

\bibitem{Van_den_Oord2018-sy}
Aaron van~den Oord, Yazhe Li, and Oriol Vinyals.
\newblock Representation learning with contrastive predictive coding.
\newblock {\em arXiv}, 1807.03748, July 2018.

\bibitem{Lowe2019-hz}
Sindy L{\"o}we, Peter O'Connor, and Bastiaan~S Veeling.
\newblock Putting an end to {End-to-End}: {Gradient-Isolated} learning of representations.
\newblock {\em arXiv}, 1905.11786, May 2019.

\bibitem{noauthor_undated-hr}
Bernd Illing, Jean Ventura, Guillaume Bellec, and Wulfram Gerstner.
\newblock Local plasticity rules can learn deep representations using self-supervised contrastive predictions.
\newblock {\em Proceedings of the 35th International Conference on Neural Information Processing Systems}, page~15, 2021.

\end{thebibliography}

\end{document}